\documentclass[12pt]{article}
\usepackage{amsmath}

\newtheorem{theorem}{Theorem}

\newtheorem{lemma}[theorem]{Lemma}
\numberwithin{equation}{section}

\newtheorem{proposition}[theorem]{Proposition}
\newtheorem{remark}[theorem]{Remark}

\input{tcilatex}

\begin{document}

\title{On H.\ Friedrich's formulation of the Einstein equations with fluid sources.}
\author{Yvonne Choquet-Bruhat and James W. York}
\maketitle

\begin{abstract}
We establish a variant of the symmetric quasi linear first order system
given by H.\ Friedrich for the evolution equations of gravitating fluid
bodies in General Relativity which can be important to solve realistic
problems. Our version has the advantage of introducing only physical
characteristics. We state explicitly the conditions under which the system
is hyperbolic and admits a well posed Cauchy problem.
\end{abstract}

\textit{Dedicated to Andrzej Granas.}

\bigskip

\section{Introduction.}

Cattaneo [C] and Ferrarese [F1], have stressed long ago that the fundamental
physical data are those defined on time lines instead of those defined on
spacelike hypersurfaces. In the case of a fluid, timelike world lines have a
material reality given by the flow, the unknowns associated to them
correspond to a lagrangian picture, in contrast with the usual eulerian one.
In many situations, particularly in cosmology, the lagrangian description
seems more fundamental. However, one cannot be too confident because General
Relativity holds many surprises. For example, consider a fluid in internal
equilibrium and in equilibrium with a stationary black hole. Under some
conditions (cf. [BY]) the thermodynamic temperature is found by certain
eulerian observers whose world lines have vanishing vorticity. This
corresponds to our ''well posedness'' result given in section 11.4. We
believe that eulerian and lagrangian descriptions are both usefull.

Ferrarese [F2], [F3] has given a complete 3+1 decomposition of the
connection and Riemann tensor in a frame whose time axis is tangent to the
time lines and the space axes orthogonal to it. Using this decomposition he
has written the Einstein- dust equations as a (infinite dimensional)
dynamical system. Recently H.\ Friedrich [HF] has used a lagangian
description of the flow together with a tetrad formalism to write the
Einstein equations coupled with a perfect fluid as a symmetric first order
system, formally hyperbolic. His treatment includes dust as a particular
case.

The Einstein- dust system, as well as the Einstein- Euler system for a
perfect fluid, have been proved long ago [C.B1] to form a well posed
hyperbolic system in the sense of Leray. The relativistic Euler system has
been put in first order symmetric hyperbolic form (FOSH) first by K.O.
Friedrichs himself [KOF] , then by general methods by Anile, Boillat and
Ruggeri (see references in [A]). Rendall [R] has obtained, for well chosen
equations of state, FOSH systems for the Euler equations which extend to
vacuum, but do not yet permit, as he points out himself, the solution of
realistic problems of motion of compact fluid bodies in otherwise empty
space. The system obtained by H.\ Friedrich can be important to solve such
realistic problems.

In this article we establish a variant of Friedrich's system for a
lagrangian and tetrad description of the flow and the spacetime geometry.
Contrary to Friedrich we will not work with the Weyl tensor but with the
Riemann tensor, as in [AC.BY]. It avoids the introduction of unphysical
characteristics and seems to us more natural in this non vacuum case as well
as somewhat simpler. The systems we obtain for vacuum, dust or perfect fluid
are symmetric and hyperbolic in the same sense as Friedrich's system. They
are hyperbolic in K.O. Friedrichs's sense, which leads to existence theorems
for solutions in Sobolev spaces, only if the time lines admit global
spacelike sections, and the initial data are given on such a section. These
existence theorems are needed even if one wants to consider the evolution
equations as a well posed dynamical system.

\section{Metric and coframe.}

We write the metric in an orthonormal frame, i.e., 
\begin{equation}
g=-(\theta ^{0})^{2}+\sum_{i=1}^{3}(\theta ^{i})^{2}.
\end{equation}
We choose the time axis to be \textbf{tangent to the time lines}, i.e. the
cobasis $\theta $ is such that $\theta ^{i}$ does not contain $dx^{0}.$ We
set 
\begin{equation}
\theta ^{i}=a_{j}^{i}dx^{j},\text{ \ \ \ \ }\theta ^{0}=Udx^{0}+b_{i}dx^{i}.
\end{equation}
We will call such a frame a C.F. (Cattaneo- Ferrarese) frame. We denote by $%
A_{i}^{j}$ the matrix inverse of $a_{i}^{j}$. It holds that 
\begin{equation*}
\text{\ \ }dx^{i}=A_{j}^{i}\theta ^{j},\text{ \ \ \ \ \ \ }%
dx^{0}=U^{-1}(\theta ^{0}-A_{j}^{i}b_{i}\theta ^{j})\text{ }.\,
\end{equation*}

The Pfaff derivatives $\partial _{\alpha }$ in the C.F. frame are linked to
the partial derivatives $\partial /\partial x^{\alpha }$ by the relations 
\begin{equation*}
\frac{\partial }{\partial x^{0}}=U\partial _{0},\text{ \ \ }\frac{\partial }{%
\partial x^{i}}=a_{i}^{j}\partial _{j}+b_{i}\partial _{0}.\,
\end{equation*}
\begin{equation*}
\partial _{0}=U^{-1}\frac{\partial }{\partial x^{0}},\text{ \ \ }\partial
_{i}=A_{i}^{j}[\frac{\partial }{\partial x^{j}}-U^{-1}b_{j}\frac{\partial }{%
\partial x^{0}}].\,
\end{equation*}

\section{ Bianchi equations}

Whatever the coframe, the components of the Riemann tensor satisfy the
identities 
\begin{equation}
\nabla _{\alpha }R_{\beta \gamma ,\lambda \mu }+\nabla _{\beta }R_{\gamma
\alpha ,\lambda \mu }+\nabla _{\gamma }R_{\alpha \beta ,\lambda \mu }\equiv
0,
\end{equation}
hence, if the Ricci tensor $R_{\alpha \beta }$ satisfies the Einstein
equations 
\begin{equation}
R_{\alpha \beta }=\rho _{\alpha \beta },
\end{equation}
it holds that 
\begin{equation}
\nabla _{\alpha }R_{\beta \gamma ::\mu ,}^{.....\alpha }=\nabla _{\beta
}\rho _{\gamma \mu }-\nabla _{\gamma }\rho _{\beta \mu }
\end{equation}
Equations 3.1 and 3.3 imply the following ones (cf. an analogous system in
[AC.BY]) 
\begin{equation}
\nabla _{0}R_{hk,0j}+\nabla _{k}R_{0h,0j}-\nabla _{h}R_{0k,0j}=0,
\end{equation}
and, in using a symmetry of the Riemann tensor, 
\begin{equation}
\nabla _{0}R_{:::i,0j}^{0}+\nabla _{h}R_{:::i,0j}^{h}=J_{i,0j}\equiv \nabla
_{0}\rho _{ji}-\nabla _{j}\rho _{0i}.
\end{equation}
The equations 3.4 and 3.5 are for each given pair ($0j)$ a first order
system for the components $R_{hk,0j\text{ }}$ and $R_{0h,0j}.$ The principal
operator is a symmetric 6 by 6 matrix:

\begin{center}
$\left( 
\begin{tabular}{cccccc}
$\partial _{0}$ & 0 & 0 & $\partial _{2}$ & -$\partial _{1}$ & 0 \\ 
0 & $\partial _{0}$ & 0 & 0 & $\partial _{3}$ & -$\partial _{2}$ \\ 
0 & 0 & $\partial _{0}$ & -$\partial _{3}$ & 0 & $\partial _{1}$ \\ 
$\partial _{2}$ & 0 & -$\partial _{3}$ & $\partial _{0}$ & 0 & 0 \\ 
-$\partial _{1}$ & $\partial _{3}$ & 0 & 0 & $\partial _{0}$ & 0 \\ 
0 & -$\partial _{2}$ & $\partial _{1}$ & 0 & 0 & $\partial _{0}$%
\end{tabular}
\right) $
\end{center}

Analogous equations and result hold with the pair $(0j)$ replaced by $(lm)$, 
$l<m$. The system finally obtained has a principal matrix consisting of 6
identical 6 by 6 blocks around the diagonal, it is symmetric.

However, the system obtained for the components of the Riemann tensor in the
coframe with time axis $\theta ^{0}$ and space axis $\theta ^{i},$ with
principal operator the matrix $M^{\alpha }\partial _{\alpha },$ cannot be
said to be hyperbolic in the usual sense: the principal matrix $M^{0}$ is
the unit matrix hence positive definite, but the operators $\partial
_{\alpha }$ are not partial derivatives. We say that the system is a quasi
FOSH (First Order Symmetric Hyperbolic) system (Cf. section 11).

\begin{remark}
One can associate with the Riemann tensor, as in the usual case of the 3+1
splitting, two pairs of ``electric'' and ``magnetic'' 2- tensors whose non
zero components in the frame are: 
\begin{align*}
E_{ij}& \equiv R_{0i,0j},\text{ \ \ }D_{ij}\equiv \frac{1}{4}\varepsilon
_{ihk}\varepsilon _{jlm}R^{hk,lm} \\
H_{ij}& \equiv \frac{1}{2}\varepsilon _{ihk}R_{:::::,0j}^{hk},\text{ \ \ }%
B_{ji}\equiv \frac{1}{2}\varepsilon _{ihk}R_{0j}^{:::::hk},
\end{align*}
where $\varepsilon _{ijk}$ is the totally antisymmetric Kronecker tensor.
These definitions imply here 
\begin{equation*}
R_{hk,0j}\equiv \varepsilon _{::hk}^{i}H_{ij},\text{ \ \ }R_{hk,lm}\equiv
\varepsilon _{::hk}^{i}\varepsilon _{::lm}^{j}D_{ij},\text{ \ \ }%
R_{0j,hk}=\varepsilon _{::hk}^{i}B_{ji}.
\end{equation*}
The previous system is a quasi FOSH system for these pairs of tensors.
\end{remark}

\begin{remark}
The system 3.1, 3.2 contains also the equations, which we do not use for
evolution, 
\begin{equation}
\nabla _{h}R_{ij,\lambda \mu }+\nabla _{j}R_{hi,\lambda \mu }+\nabla
_{i}R_{jh,\lambda \mu }=0
\end{equation}
and 
\begin{equation}
\nabla _{h}R_{..0,\lambda \mu }^{h}=\nabla _{\lambda }\rho _{\mu 0}-\nabla
_{\mu }\rho _{\lambda 0}.
\end{equation}
These equations are not usual constraints on a submanifold $t=cons\tan t$
because $\partial _{i}$ contains the transversal derivative $\partial
/\partial t.$ We call them Bianchi quasi-constraints.
\end{remark}

\section{Coframe structure coefficients.}

The structure coefficients $c$ of the coframe are defined by 
\begin{equation*}
d\theta ^{\alpha }\equiv -\frac{1}{2}c_{\beta \gamma }^{\alpha }\theta
^{\beta }\wedge \theta ^{\gamma }.
\end{equation*}
We have 
\begin{equation*}
d\theta ^{0}=\partial _{i}U\theta ^{i}\wedge dx^{0}+\partial _{\alpha
}b\theta ^{\alpha }\wedge dx^{i}
\end{equation*}
that is, 
\begin{equation*}
d\theta ^{0}=U^{-1}\partial _{i}U\theta ^{i}\wedge (\theta
^{0}-A_{j}^{h}b_{h}\theta ^{j})+A_{j}^{h}\partial _{\alpha }b_{h}\theta
^{\alpha }\wedge \theta ^{j},
\end{equation*}
therefore 
\begin{equation*}
c_{0i}^{0}=U^{-1}\partial _{i}U-A_{i}^{j}\partial _{0}b_{j}=-c_{i0}^{0}
\end{equation*}
and, with $f_{[ij]}\equiv f_{ij}-f_{ji},$%
\begin{equation*}
c_{ij}^{0}=U^{-1}b_{h}A_{[j}^{h}\partial _{i]}U-A_{[j}^{h}\partial _{i]}b_{h}
\end{equation*}
that is 
\begin{equation*}
c_{ij}^{0}=UA_{[i}^{h}\partial _{j]}(U^{-1}b_{h})
\end{equation*}
also 
\begin{equation*}
d\theta ^{i}=A_{k}^{j}(\partial _{h}a_{j}^{i}\theta ^{h}\wedge \theta
^{k}+\partial _{O}a_{j}^{i}\theta ^{0}\wedge \theta ^{k})
\end{equation*}
hence 
\begin{equation*}
c_{k0}^{i}=A_{k}^{j}\partial _{0}a_{j}^{i}
\end{equation*}
\begin{equation*}
c_{hk}^{i}=A_{[h}^{j}\partial _{k]}a_{j}^{i}
\end{equation*}

\begin{remark}
If the time lines are not hypersurface orthogonal (i.e. if $b_{i}\not =0)$
the coefficients $c_{ij}^{h}$ are different from the structure coefficients
of the space frame $\theta^{i}.$
\end{remark}

\section{Connection.}

The connection is such that 
\begin{equation*}
\omega _{\beta \gamma }^{\alpha }-\omega _{\gamma \beta }^{\alpha }=c_{\beta
\gamma }^{\alpha }
\end{equation*}
and 
\begin{equation*}
\omega _{\gamma \beta ,\lambda }+\omega _{\gamma \lambda ,\beta }=0
\end{equation*}
with ($\eta $ is the Minkowski metric) 
\begin{equation*}
\omega _{\gamma \beta ,\lambda }\equiv \eta _{\alpha \lambda }\omega
_{\gamma \beta }^{\alpha }
\end{equation*}
We set 
\begin{equation*}
c_{\gamma \beta ,\lambda }\equiv \eta _{\alpha \lambda }c_{\gamma \beta
}^{\alpha },
\end{equation*}
and we have (Cf. C.B-DeWitt I p.308$)$%
\begin{equation*}
\omega _{\alpha \lambda ,\mu }=\frac{1}{2}(c_{\alpha \lambda ,\mu
}-c_{\alpha \mu ,\lambda }-c_{\lambda \mu ,\alpha }).
\end{equation*}
We find that, as foreseen from antisymmetry 
\begin{equation*}
\omega _{00}^{0}=\omega _{i0}^{0}
\end{equation*}
We set 
\begin{equation*}
Y_{i}\equiv \omega _{00,i},
\end{equation*}
and we have 
\begin{equation*}
Y_{i}\equiv \omega _{00,i}=\omega _{0i}^{0}=\omega
_{00}^{i}=-c_{0i,0}=c_{0i}^{0}=U^{-1}\partial _{i}U-A_{i}^{j}\partial
_{0}b_{j}.
\end{equation*}
We know that $\omega _{0i}^{j}=\omega _{0i,j}$ is antisymmetric in $i$ and $%
j.$ We set 
\begin{equation*}
\omega _{0i,j}\equiv f_{ij}=\frac{1}{2}\{A_{j}^{h}\partial
_{0}a_{h}^{i}-A_{i}^{h}\partial _{0}a_{i}^{j}+A_{[i}^{h}\partial
_{j]}b_{h}\}.
\end{equation*}

\begin{remark}
Let $e_{(\alpha )}\equiv \partial _{\alpha }$ $\ $be the frame dual to $%
\theta ^{\alpha }$, i.e. with components $\delta _{(\alpha )}^{\lambda }.$
Then 
\begin{equation*}
\nabla _{\beta }e_{(\alpha )}^{\lambda }=\omega _{\beta \alpha }^{\lambda },
\end{equation*}
in particular $\omega _{0i,j}$ is the projection on $e_{(j)}$ of the
derivative of $e_{(i)}$ in the direction of $e_{(0)}.$ We have $f_{ij}=0$ if
the frame is Fermi transported along the time line.
\end{remark}

The connection coefficient $\omega _{i0,j}$ is the sum of a term symmetric
in $i$ and $j$ and an antisymmetric one, we have 
\begin{equation*}
X_{ij}\equiv \omega _{i0,j}=\omega _{i0}^{j}=\frac{1}{2}\{A_{j}^{h}\partial
_{0}a_{h}^{i}+A_{i}^{h}\partial _{0}a_{i}^{j}+A_{[i}^{h}\partial
_{j]}b_{h}\}.
\end{equation*}
The antisymmetric term vanishes if the time lines are hypersurface
orthogonal ($b_{i}=0).$

The coefficients $\omega_{ij}^{h}$ are also linear expressions in terms of
the first derivatives of the frame coefficients, they are identical to the
connection constructed with the $a_{i}^{j}$ at fixed $x^{0}$ if $b_{i}=0.$

\section{Frame evolution.}

The relations between the connection coefficients $\omega _{ij}^{h}$ and the
frame coefficients $a_{i}^{j}$ give equations for the dragging of these
coefficients along the time lines when the connection is known. Indeed we
have found 
\begin{equation}
\partial _{0}a_{j}^{i}=a_{j}^{k}c_{0k}^{i}=2a_{j}^{k}(\omega
_{0k}^{i}-\omega _{k0}^{i})\equiv 2a_{j}^{k}(f_{k}^{i}-X_{k}^{i})
\end{equation}
\begin{equation}
\partial _{0}b_{i}=a_{i}^{h}(-Y_{h}+U^{-1}\partial _{h}U
\end{equation}

No equation gives the evolution of $U$. It can be considered as a gauge
variable fixing the time parameter. H.\ Friedrich chooses the coordinate $%
x^{0}$ to be the proper time along the given time lines, i.e. $U=1$

The quantity $f_{j}^{i}$ is also a gauge variable fixing the evolution of
the space frame. If we choose it to be Fermi transported like Friedrich,
then $f_{j}^{i}=0.$

\section{Curvature.}

\subsection{Evolution of connection.}

Using the general formula (cf. C.B- DeWitt I p. 306, with opposite sign
convention) 
\begin{equation}
R_{\lambda \mu ....\beta }^{......\alpha }\equiv \partial _{\lambda }\omega
_{\mu \beta }^{\alpha }-\partial _{\mu }\omega _{\lambda \beta }^{\alpha
}-\omega _{\rho \beta }^{\alpha }(\omega _{\lambda \mu }^{\rho }-\omega
_{\mu \lambda }^{\rho })+\omega _{\lambda \rho }^{\alpha }\omega _{\mu \beta
}^{\rho }-\omega _{\mu \rho }^{\alpha }\omega _{\lambda \beta }^{\rho }
\end{equation}
\begin{equation}
R_{0h..j}^{.....i}\equiv \partial _{0}\omega _{hj}^{i}-\tilde{\nabla}%
_{h}f_{j}^{i}-Y_{h}f_{j}^{i}-(f_{h}^{k}-X_{h}^{..k})\omega
_{kj}^{i}+Y^{i}X_{hj}^{..}-Y_{j}X_{h}{}^{i}
\end{equation}
where $\tilde{\nabla}$ is the pseudo covariant derivative constructed with $%
\partial _{i}$ and $\omega _{ij}^{h}$ (Cattaneo- Ferrarese transversal
derivative). In the Fermi gauge $R_{0h..j}^{.....i}$ reduces to 
\begin{equation}
R_{0h..j}^{.....i}\equiv \partial _{0}\omega _{hj}^{i}+X_{h}^{..k}\omega
_{kj}^{i}+Y_{i}X_{h}^{...j}-Y_{j}X_{h}{}^{i}.
\end{equation}
The principal operator in these equations is the dragging along the time
lines of $\omega _{ij}^{h}.$

We have: 
\begin{equation}
R_{h0..0}^{.....i}\equiv \tilde{\nabla}_{h}Y_{i}-\partial
_{0}X_{h}^{..i}+Y^{i}Y_{h}-X_{h}^{..j}X_{j}^{..i}+f_{h}^{..j}X_{j}^{..i}-f_{j}^{..i}X_{h}^{..j}.
\end{equation}
We deduce from this identity, in the Fermi gauge, 
\begin{equation}
R_{00}\equiv \tilde{\nabla}_{i}Y^{i}-\partial
_{0}X_{i}^{i}+Y^{i}Y_{i}-X_{h}^{..j}X_{j}^{..h}.
\end{equation}

In the next sections we will write other equations to achieve the
determination of both $Y$ and $X.$

\subsection{Quasi-constraints.}

The other components of the Riemann tensor are given in the C-F frame by:

\begin{equation}
R_{hk..j}^{.....i}\equiv \tilde{R}%
_{hk...j}^{......^{i}}-f_{j}^{i}(X_{hk}-X_{kh})+X_{.k}^{.i}X_{jh}-X_{jk}^{.}X_{..h}^{i},
\end{equation}
where $\tilde{R}_{hk..j}^{.....^{i}}$ denotes the expression formally
constructed as a Riemann tensor with the coefficients $\omega _{ij}^{h}.$ 
\begin{equation}
R_{kh....j}^{......0}\equiv \partial _{k}\omega _{hj}^{0}-\partial
_{h}\omega _{kj}^{0}-\omega _{\rho j}^{0}(\omega _{kh}^{\rho }-\omega
_{hk}^{\rho })+\omega _{k\rho }^{0}\omega _{hj}^{\rho }-\omega _{h\rho
}^{0}\omega _{kj}^{\rho },
\end{equation}
that is ,with previous notations, 
\begin{equation}
R_{kh....j}^{......0}\equiv \tilde{\nabla}_{k}X_{hj}-\tilde{\nabla}%
_{h}X_{kj}-Y_{j}(X_{kh}-X_{hk}).
\end{equation}
These equations do not enter in the evolution system for the connection that
we are considering, they do not contain the derivative $\partial _{0},$ we
call them connection quasi- constraints.

\begin{remark}
Modulo the expression of the connection in terms of frame coefficients one
has the well known symmetry: 
\begin{equation*}
R_{kh,0j}=R_{0j,kh}
\end{equation*}
\end{remark}

One deduces from the identity 7.8 
\begin{equation*}
R_{h0}\equiv\tilde{\nabla}_{j}X_{h}^{...j}-\tilde{\nabla}%
_{h}X_{j}^{j}-Y^{j}(X_{jh}-X_{hj})
\end{equation*}

\section{Vacuum case.}

In the vacuum case there are no a-priori given time lines. We can give
arbitrarily on the spacetime the projection $Y_{i}$ of $\nabla
_{e_{(0)}}e_{(0)}$ on $e_{(i)}.$ The quantities $U$ and $f_{i}^{j}$ being
also arbitrarily given the Bianchi equations together with 6-1, 6-2; 7-3;
7-5 constitute a quasi FOSH system for the unknowns $E_{ij},$ $D_{ij},$ $%
H_{ij},$ $B_{ij},$ $a_{i}^{j},$ $b_{i},$ $\omega _{ij}^{h},$ $X_{ij}.$ Its
solution determines the spacetime metric.

\section{Perfect fluid.}

\subsection{Fluid equations.}

The stress energy tensor of a perfect fluid is 
\begin{equation*}
T_{\alpha \beta }=(\mu +p)u_{\alpha }u_{\beta }+pg_{\alpha \beta }.
\end{equation*}
Then 
\begin{equation*}
\rho _{\alpha \beta }=(\mu +p)u_{\alpha }u_{\beta }+\frac{1}{2}(\mu
-p)g_{\alpha \beta }.
\end{equation*}
One supposes that the matter energy density $\mu $ is a given function of
the pressure $p$ and entropy $S;$ this last function is conserved along the
flow lines 
\begin{equation}
u^{\alpha }\partial _{\alpha }S=0.
\end{equation}
The Euler equations of the fluid express the generalized conservation law 
\begin{equation*}
\nabla _{\alpha }T^{\alpha \beta }=0,
\end{equation*}
and are equivalent to the equations 
\begin{equation*}
(\mu +p)u^{\alpha }\nabla _{\alpha }u^{\beta }+(u^{\alpha }u^{\beta
}+g^{\alpha \beta })\partial _{\alpha }p=0,\text{ \ \ with \ \ \ }u^{\alpha
}u_{\alpha }=-1
\end{equation*}
and 
\begin{equation*}
(\mu +p)\nabla _{\alpha }u^{\alpha }+u^{\alpha }\partial _{\alpha }\mu =0
\end{equation*}
In our coframe they reduce to: 
\begin{equation}
\partial _{0}S=0
\end{equation}
\begin{equation}
(\mu +p)Y_{i}+\partial _{i}p=0,\text{ \ \ \ }Y_{i}\equiv \omega _{00}^{i}
\end{equation}
\begin{equation}
\partial _{0}\mu +(\mu +p)X_{i}^{i}=0.
\end{equation}
Using the index $F$ of the fluid defined by 
\begin{equation*}
F(p,S)=\int \frac{dp}{\mu (p,S)+p},
\end{equation*}
the equation 9.3 reads 
\begin{equation}
Y_{i}=-\partial _{i}F,
\end{equation}
while 9.4 gives , modulo 9.2, 
\begin{equation}
\mu _{p}^{\prime }\partial _{0}F+X_{i}^{i}=0.
\end{equation}
The commutation relation between Pfaff derivatives 
\begin{equation*}
(\partial _{0}\partial _{i}-\partial _{i}\partial _{0})F=c_{0i}^{\alpha
}\partial _{\alpha }F
\end{equation*}
implies therefore 
\begin{equation}
\mu _{p}^{\prime }[\partial _{0}Y_{i}+Y_{i}\partial
_{0}F+(f_{i}^{j}-X_{i}^{j})\partial _{j}F]-\partial _{i}\mu _{p}^{\prime
}\partial _{0}F-\partial _{i}X_{h}^{h}=0.
\end{equation}
The use of 9.5 and 9.6 replaces $\partial _{\alpha }F$ by functions of $Y,X$
and $p,S.$ The derivatives $\partial _{i}\mu _{p}^{\prime }$ are functions
of $Y,p,S$ and $\partial _{k}S,$ since 
\begin{equation*}
\partial _{i}\mu _{p}^{\prime }=\mu "_{p^{2}}\partial _{i}p+\mu
"_{pS}\partial _{i}S.
\end{equation*}
We introduce $S_{k}=\partial _{k}S$ as new unknowns. They are dragged along
the flow lines by the following equation deduced from 9.2: 
\begin{equation}
\partial _{0}S_{k}=c_{0k}^{j}S_{j}\equiv (f_{k}^{j}-X_{j}^{k})S_{j}.
\end{equation}

Following H.\ Friedrich we replace in 9.7 $\partial _{i}X_{h}^{h}$ by its
expression deduced from the equation 
\begin{equation*}
R_{i0}\equiv \tilde{\nabla}_{h}X_{i}^{...h}-\tilde{\nabla}%
_{i}X_{h}^{h}-Y^{h}(X_{hi}-X_{ih})=0,
\end{equation*}
and we obtain, changing names of indices 
\begin{equation*}
\mu _{p}^{\prime }\partial _{0}Y_{h}-\tilde{\nabla}%
_{j}X_{h}^{j}-Y^{h}(X_{hi}-X_{ih})+\mu _{p}^{\prime }[Y_{h}\partial _{0}F+
\end{equation*}
\begin{equation}
(f_{h}^{j}-X_{h}^{j})\partial _{j}F]+\partial _{h}\mu _{p}^{\prime }\partial
_{0}F=0.
\end{equation}
In 7.4 we replace $\tilde{\nabla}_{h}Y_{i}$ by $\tilde{\nabla}%
_{i}Y_{h}+c_{hi}^{0}Y_{0},$ with 
\begin{equation*}
Y_{0}\equiv -\partial _{0}F\equiv -(\mu _{p}^{\prime })^{-1}X_{i}^{i},\text{
\ \ }c_{hi}^{0}\equiv X_{ih}-X_{hi}
\end{equation*}
and we obtain 
\begin{equation}
\partial _{0}X_{h}^{.i}-\tilde{\nabla}%
^{i}Y_{h}-c_{hi}^{0}Y_{0}-Y_{h}Y^{i}+(X_{h}^{..j}-f_{h}^{..j})X_{j}^{..i}+f_{j}^{..i}X_{h}^{..j}=-R_{h0..0}^{.....i}.
\end{equation}
The principal operator on the unknowns $Y$ and $X$ in the equations 9.9,
9.10 is diagonal by blocks. Each block, corresponding to a given index $h,$
is symmetric, it reads:

\begin{center}
$\left( 
\begin{tabular}{cccc}
$\mu _{p}^{\prime }\partial _{0}$ & -$\partial _{1}$ & -$\partial _{2}$ & -$%
\partial _{3}$ \\ 
-$\partial _{1}$ & $\partial _{0}$ & 0 & 0 \\ 
-$\partial _{2}$ & 0 & $\partial _{0}$ & 0 \\ 
-$\partial _{3}$ & 0 & 0 & $\partial _{0}$%
\end{tabular}
\right) $
\end{center}

If $\mu _{p}^{\prime }>0$ the matrix $M^{0}$ is positive definite in the
C.F. frame. The system 9.9, 9.10 is a quasi FOSH system for the pairs $%
Y_{h},X_{h}^{j}.$

\begin{remark}
The characteristic determinant associated with the system 9.9, 9.10,
obtained by replacing $\partial _{\alpha }$ with a covariant vector $\xi $
is 
\begin{equation*}
\{\xi _{0}^{2}(\mu _{p}^{\prime }\xi _{0}^{2}-\sum_{i=1,2,3}\xi
_{i}^{2})\}^{3}
\end{equation*}
The roots of $\mu _{p}^{\prime }\xi _{0}^{2}-\sum_{i=1,2,3}\xi _{i}^{2}=0$
correspond to sound waves. Their speed is at most 1 (speed of light) if and
only if $\mu _{p}^{\prime }\geq 1.$
\end{remark}

The full system of fluid equations is quasi diagonal by blocks. The matrices 
$M^{\alpha }\partial _{\alpha }$ corresponding to 9.2, 9.4, 9.8 reduce to $%
\partial _{0}.$

\subsection{Sources of the Bianchi equations.}

In our frame the source tensor $\rho _{\alpha \beta }$ reduces to 
\begin{equation*}
\rho _{00}=\frac{1}{2}(\mu +3p),\text{ \ \ }\rho _{0i}=0,\text{ \ \ }\rho
_{ij}=\frac{1}{2}(\mu -p)\delta _{ij}
\end{equation*}
We have seen that $\partial _{\alpha }p$ and $\partial _{\alpha }\mu $ are
smooth functions of $p,S,S_{i},Y$ and $X.$ The same property holds for $%
J_{i,0j}$ and $J_{i,hk}.$

\subsection{Conclusion.}

Assembling the results of the previous sections we find the following
theorem.

\begin{theorem}
The EEF (Einstein-Euler-Friedrich) system for a gravitating perfect fluid,
with flow lines taken as time lines, $U$ and $f_{j}^{i}\equiv \omega
_{0j}^{i}$ given arbitrarily, is a quasi FOSH system for the Riemann
curvature tensor, the frame and connection coefficients, the density of
matter, the entropy and its space derivatives.
\end{theorem}

As remarked in section 3 the system is not a usual FOSH system: the time
lines are by choice timelike but the hypersurfaces $x^{0}=cons\tan t$ are
not necessarily spacelike for the characteristic cone.

\section{Case of dust.}

\subsection{Matter equations.}

The stress energy tensor of a dust source is 
\begin{equation*}
T_{\alpha \beta }\equiv \mu u_{\alpha }u_{\beta }.
\end{equation*}
The flow lines, tangent to the unit vector $u^{\alpha },$ are then
geodesics, it holds that 
\begin{equation}
u^{\alpha }\nabla _{\alpha }u^{\beta }=0,
\end{equation}
while the conservation of matter reads: 
\begin{equation}
u^{\alpha }\partial _{\alpha }\mu +\mu \nabla _{\alpha }u^{\alpha }=0.
\end{equation}

We take the flow lines as time lines, i.e. $u^{\alpha }=\delta _{0}^{\alpha
}.$ Then 10.2 reads 
\begin{equation}
\partial _{0}\mu +\mu X_{i}^{i}=0,
\end{equation}
while 10-1 gives 
\begin{equation}
\omega _{00}^{i}\equiv Y_{i}=0.
\end{equation}
Using this equation we see that, given arbitrarily on the spacetime the
gauge variables $a_{0}^{0}$ the equations (6-1), (6-2) reduce to a quasi-
diagonal system with principal operator $\partial _{0}$ for the frame
coefficients when the connection (which appears undifferentiated) is known .

\subsection{Bianchi equations.}

For a dust stress energy tensor it holds that 
\begin{equation*}
\rho _{\alpha \beta }=\mu (u_{\alpha }u_{\beta }+\frac{1}{2}g_{\alpha \beta
}),
\end{equation*}
hence in the chosen frame 
\begin{equation*}
\rho _{00}=\frac{1}{2}\mu ,\text{ \ \ }\rho _{0i}=0,\text{ \ \ }\rho _{ij}=%
\frac{1}{2}\mu \delta _{ij}
\end{equation*}
A straightforward computation gives then, choosing the Fermi transported
frame for simplicity 
\begin{equation*}
J_{i,0j}\equiv \nabla _{0}\rho _{ji}-\nabla _{j}\rho _{0i}=\mu (-\frac{1}{2}%
\delta _{ij}X_{h}^{h}+X_{ji})
\end{equation*}
\begin{equation*}
J_{i,hj}\equiv \nabla _{h}\rho _{ji}-\nabla _{j}\rho _{hi}=\frac{1}{2}%
\{\partial _{h}\mu \delta _{ji}-\partial _{j}\mu \delta _{hi}+\mu (\omega
_{hj}^{i}-\omega _{jh}^{i})\}
\end{equation*}

There appears a difficulty which is the presence of the space derivatives $%
\partial _{h}\mu $ in $J_{i,hj}.$ We get rid of this problem, inspired
(though with a different choice) by H. Friedrich's treatment, by replacing
the unknown $R_{hk..j}^{.....i}$ by another unknown with the same
symmetries, namely 
\begin{equation*}
F_{ij,hk.}=R_{ij,hk}+\delta _{ik}R_{jh}-\delta _{ih}R_{jk}-\delta
_{jk}R_{ih}+\delta _{ih}R_{ik}+D_{ij,hk}R
\end{equation*}
with 
\begin{equation*}
D_{ij,hk}\equiv \frac{1}{2}(\delta _{ik}\delta _{jh}-\delta _{jk}\delta
_{ih})
\end{equation*}
Using this unknown, the contracted Bianchi identities become 
\begin{equation}
\nabla _{0}R_{:::i,hj}^{0}+\nabla _{k}F_{:::i,hj}^{k}\equiv 0,
\end{equation}
while the original ones become 
\begin{equation*}
\nabla _{0}F_{ij,hk}+\nabla _{j}R_{0i,hk}-\nabla _{i}R_{0j,hk}\equiv
\end{equation*}
\begin{equation}
\nabla _{0}\{\delta _{ik}R_{jh}-\delta _{ih}R_{jk}-\delta _{jk}R_{ih}+\delta
_{ih}R_{ik}+X_{ijhk}R\}
\end{equation}
One obtains a quasi FOSH system for $R_{0i,hk}$ and $F_{ij,hk}$ by replacing
on the right hand side of this equation $R_{\alpha \beta }$ by $\rho
_{\alpha \beta }$: now there appears only the time derivative $\partial
_{0}\mu $ which, satisfying 10-3, can be eliminated in favor of
undifferentiated terms.

Note that $R_{ij,hk}$ does not appear in other equations than the Bianchi
equations.

\subsection{Conclusion.}

We have proved the following theorem

\begin{theorem}
The EDFF (Einstein-Dust-Ferrarese-Friedrich) system, , with flow lines taken
as time lines, $U$ and $f_{j}^{i}\equiv \omega _{0j}^{i}$ given arbitrarily,
is a quasi FOSH system for the Riemann curvature tensor, the frame and
connection coefficients, the density of matter and its space derivatives.
\end{theorem}

\section{Cauchy problem.}

\subsection{Hyperbolicity.}

The fact that for the systems that we have obtained the hypersurfaces $%
x^{0}=cons\tan t$ are not necessarily spacelike for the characteristic cones
poses a difficulty for the Cauchy problem, since energy estimates used in
proving existence theorems rely on this property.

We have, using standard definitions

\begin{proposition}
A quasi FOSH system, with principal operator the matrix $M^{\alpha }\partial
_{\alpha },$ is a FOSH system with $x^{0}$ as a time variable if the matrix
of coefficients of $\partial /\partial x^{0}$ is positive definite, namely
if 
\begin{equation*}
\tilde{M}^{0}\equiv M^{\alpha }A_{\alpha }^{0}\equiv
U^{-1}(M^{0}-A_{j}^{i}b_{i}M^{j})
\end{equation*}
is positive definite.
\end{proposition}

\subsubsection{Case of dust}

\begin{lemma}
The EDFF (Einstein-Dust-Ferrarese-Friedrich) system is a FOSH system
relative to $x^{0}=cons\tan t$ \ slices as long as the quadratic form 
\begin{equation}
\bar{g}_{jh}=\sum_{i=1,2,3}a_{j}^{i}a_{h}^{i}-b_{j}b_{h}
\end{equation}
is positive definite and $U>0.$
\end{lemma}

Proof.

The matrix $\tilde{M}_{(bian)}^{0}$ of coefficients of $\partial /\partial
x^{0}$ corresponding to the Bianchi equations is after multiplication by $U$%
, setting 
\begin{equation*}
B_{i}\equiv -A_{i}^{j}b_{i},
\end{equation*}

\begin{center}
$\left( 
\begin{array}{cccccc}
1 & 0 & 0 & B_{2} & -B_{1} & 0 \\ 
0 & 1 & 0 & 0 & B_{3} & -B_{2} \\ 
0 & 0 & 1 & -A_{3} & 0 & B_{1} \\ 
B_{2} & 0 & -B_{3} & 1 & 0 & 0 \\ 
-B_{1} & B_{3} & 0 & 0 & 1 & 0 \\ 
0 & -B_{2} & B_{1} & 0 & 0 & 1
\end{array}
\right) $
\end{center}

Its eigenvalues are, each with multiplicity 2: 
\begin{equation*}
1,1+\sqrt{\left( B_{2}^{2}+B_{3}^{2}+B_{1}^{2}\right) },1-\sqrt{\left(
B_{2}^{2}+B_{3}^{2}+B_{1}^{2}\right) }
\end{equation*}
They are positive, because the given condition implies $%
B_{2}^{2}+B_{3}^{2}+B_{1}^{2}<1.$

The other matrices $\tilde{M}_{(conn)}^{0}$ and $\tilde{M}_{(matter)}^{0}$
are unit matrices.

\subsubsection{Fluid case.}

\begin{lemma}
The EEF (Einstein-Euler-Friedrich) system is a FOSH system relatively to $%
x^{0}=cons^{t}$ slices as long as the quadratic form $\bar{g}_{jh}$ given in
11.1 is positive definite, $U>0$ and $\mu _{p}^{\prime }\geq 1.$
\end{lemma}

Proof. The principal operator is symmetric and composed of blocks around the
diagonal.

The matrix $\tilde{M}_{(bian)}^{0}$ is the same as in the case of dust.

The matrix $\tilde{M}_{con}^{0}$ corresponding to the connection evolution
is, after multiplication by $U$

\begin{center}
$\left( 
\begin{array}{cccc}
\mu _{p}^{\prime } & -B_{1} & -B_{2} & -B_{3} \\ 
-B_{1} & 1 & 0 & 0 \\ 
-B_{2} & 0 & 1 & 0 \\ 
-B_{3} & 0 & 0 & 1
\end{array}
\right) .$
\end{center}

Its eigenvalues are $1,$ with multiplicity two, and: 
\begin{equation}
\frac{1}{2}\{1+\mu _{p}^{\prime }\pm 2\sqrt{\frac{1}{4}\left( \mu
_{p}^{\prime }-1)^{2}+\sum B_{i}^{2}\right) }\}
\end{equation}
These eigenvalues are positive under the hypothesis 11-1 if $\mu
_{p}^{\prime }\geq 1.$

\begin{remark}
The condition $\mu_{p}^{\prime}\geq1$ expresses that the \ sound speed is at
most equal to the speed of light.
\end{remark}

\subsection{Cauchy data.}

The Cauchy problem is less natural in the C.F. formulation than in the usual
3+1 decomposition.

\subsubsection{Initial data.}

An initial data set on a coordinate patch of a 3 dimensional submanifold is
composed of the following elements:

1. A field of coframes $\bar{a}_{j}^{i}{}^{j}$and a field of covariant
vectors \ $\bar{a}_{i}^{0}$

The quadratic form on $M$ with coefficients 
\begin{equation}
\bar{g}_{jh}=\sum_{i=1,2,3}\bar{a}_{j}^{i}\bar{a}_{h}^{i}-\bar{b}_{j}\bar{b}%
_{h}
\end{equation}
is supposed to be positive definite. The submanifold $M_{0}\equiv M\times
\{0\}$ of $M\times R$ is then spacelike for any lorentzian metric on $%
M\times R$ which reduces on $M$ to 
\begin{equation*}
-(\bar{U}dx^{0}+\bar{b}_{i}dx^{i})^{2}+\sum_{i=1,2,3}(\bar{\theta}^{i})^{2},%
\text{ \ \ }\bar{U}>0
\end{equation*}

2. Quantities $\bar{\omega}_{ij}^{h},$ $\bar{X}_{i}^{j},\bar{Y}_{i}.$

3. Pairs of 2-tensors $\bar{E}^{ij},\bar{H}_{ij},\bar{D}_{ij},\bar{B}_{ij}.$

4.a (case of dust) A scalar function $\bar{\mu}\geq0.$

4.b (perfect fluid case). Two scalar functions $\bar{p}\geq0$ and $\bar{S}%
\geq0$, and equation of state $\mu(p,S)$ such that $\mu_{p}^{\prime}\geq1$
for all $p$ in a neighbourhood of $(\bar{p},\bar{S})$.

\subsubsection{Constraints.}

The initial data on $M$ do not determine on $M$ the derivatives $\partial
_{i}\equiv \bar{\partial}_{i}-A_{i}^{j}\bar{b}_{j}\partial _{0}.$ To satisfy
the quasi-constraints on $M$ we must first deduce from the given data and
the considered evolution system the values on $M$ of the derivatives with
respect to $x^{0}$ of the relevant unknowns. The quasiconstraints give then
relations on the initial data which we call constraints and suppose to be
satisfied on $M.$

\subsection{Existence theorem for the reduced system.}

Local existence theorems for a solution of the Cauchy problem are a direct
consequence of known results for quasilinear FOSH systems. We enunciate the
result in the perfect fluid case.\ The spaces $H_{s}^{u.loc}(M)$ are the
usual uniformly local Sobolev spaces on $M.$

\begin{theorem}
Let there be given an initial data set in $H_{s}^{u.loc}(M),s>\frac{3}{2}+1.$
Then for any choice on $M\times I$ of the gauge variables in $%
C^{0}(I,H_{s}^{u.loc}(M)),$with $I$ an interval of $R$ containing $0,$ there
exists a neighbourhood $I^{\prime }$ of $0$ in $R$ such that the EEF
equations admit a solution in $C^{0}(I^{\prime },H_{s}^{u.loc}(M))\cap
C^{1}(I^{\prime },H_{s-1}^{u.loc}(M)).$
\end{theorem}

To prove that the solution satisfies, modulo the initial constraints, the
original Einstein- Euler equations would be a messy task, which we will not
endeavour.\ The uniqueness of the obtained solution gives an indirect proof,
modulo the existence theorem already known [C.B1], [C.B2].

\subsection{Irrotational flows.}

We recall the equation 6-2: 
\begin{equation}
\partial _{0}b_{i}=a_{i}^{h}(-Y_{h}+\partial _{h}\log U),
\end{equation}
\ with, in the case of a perfect fluid,\ \ 
\begin{equation*}
Y_{h}=-\partial _{h}F.
\end{equation*}
We choose then (gauge choice, the geometrical result is independent of it) 
\begin{equation*}
U=\exp (-F).
\end{equation*}
The equation reduces then to 
\begin{equation}
\partial _{0}b_{i}=0\text{.}
\end{equation}
Therefore it holds that \ $b_{i}=0$ througout the motion if it is so
initially. It is the well known property of conservation of zero vorticity
(irrotationality) of a perfect fluid flow.

For an irrotational flow it holds that $X_{ij}$ is a symmetric 2- tensor,
equal up to the sign to the extrinsic curvature of the space slices $%
x^{0}=cons\tan t$ (components in the frame $\theta ^{i})$, and $%
Y_{i}=U^{-1}\partial _{i}U,$ \ while $\omega _{ij}^{h}=\bar{\omega}_{ij}^{h}$
are the connection coefficients of the space metric $\bar{g}%
=\sum_{i=1,2,3}(\theta ^{i})^{2},$ $\theta ^{i}\equiv a_{h}^{i}dx^{h}.$ The
pseudo covariant derivatives $\tilde{\nabla}_{i}$ are covariant derivatives $%
\bar{\nabla}_{i}$ in the space metric.

For an irrotational flow the EFF system is a FOSH system.

The same result holds in the case of dust the appropriate gauge choice is
then $U=cons\tan t.$

\begin{remark}
The relations between frame and connection, as well as connection and
curvature, for hypersurface orthogonal time lines are a special case of
those found in the 3+1 decomposition [C.B3]. In particular, in the present
notations, the Einstein constraints take the familiar form: 
\begin{equation}
R_{i0}\equiv \bar{\nabla}_{j}X_{i}^{j}-\partial _{i}X_{j}^{j}=0
\end{equation}
\begin{equation}
S_{00}\equiv \frac{1}{2}(R_{00}+\delta ^{ij}R_{ij})\equiv \frac{1}{2}(\bar{R}%
-X_{i}^{.j}X_{j}^{.i}+(X_{i}^{i})^{2})=\rho _{00}=\frac{1}{2}(\mu +3p).
\end{equation}
\end{remark}

\textbf{Bibliography.}

[A] Anile M. ''Relativistic fluids and magneto- fluids'' (1989) Cambridge
Univ. Press.

[AC.BY] Anderson A., Choquet-Bruhat Y., York, J.W. (1997) ''Einstein-
Bianchi hyperbolic system for General Relativity'' Top. Meth. non lin. anal.
10, 353-373.

[BY] Brown D., York J. (1994) Physical origins of time asymmetry, Halliwell,
Perez- Mercador, Zurek eds. Cambridge Univ. Press 465-474.

[C] Cattaneo C. (1959) Proiezioni naturali e derivate trasversa in una
varieta a metrica iperbolica normale Ann. Math. pura e app. (IV) XLVIII ,
361-386.

[C.B1] (Choquet) Foures -Bruhat Y.\ (1958) ''Th\'{e}or\`{e}mes d'existence
en m\'{e}canique des fluides relativistes'' Bull. Soc. Math. France, 86
155-175.

[C.B2] (Choquet) -Bruhat Y. (1966) ''Etude des \'{e}quations des fluides
charg\'{e}s relativistes inductifs et conducteurs'' Comm. Math. Phys. 3,
334-357.

[C.B3] (Choquet) Foures -Bruhat Y.\ (1956) ''Sur l'int\'{e}gration des
\'{e}quations de la Relativit\'{e} G\'{e}n\'{e}rale'' J. Rat. Mech. and
Anal. 5, 951-966.

[F1] Ferrarese G.\ (1963) ''Contributi a la tecnica delle proiezioni in una
variet\`{a} a metrica iperbolica normale'' Rendic. matem. 22, 147-168.

[F2] Ferrarese G.\ (1979) Introduzione a la dinamica riemanniana dei sistemi
continui, Ed. pitagora, Bologna.

[F3] Ferrarese G. and Catani C.\ (1994) ''Generalized frames of references
and intrinsic Cauchy problem in General Relativity'' in ''Physics on
manifolds'', Flato M., Kerner R., Lichnerowicz A. ed. Reidel, 93-108.

[HF] Friedrich H. (1998) ''On the evolution equations for gravitating ideal
fluid bodies in General Relativity'' Phys. Rev. D 57, 2317-2322.

[KOF] Friedrichs K.\ O., (1974) ''On the laws of relativistic electro-
magneto- fluid dynamics Comm. pure app. math. XXVII, 749-808.

[R] Rendall A.\ (1992) ''The initial value problem for a class of general
relativistic fluid bodies'' J.\ Math. Phys. 33, 1047-1053.

Yvonne Choquet-Bruhat, Universit\'{e} Paris 6, tour 22-12, 4 place Jussieu
75230 Paris, France.

James W. York, Department of Physics, UNC, Chapel Hill, N.C, U.S.A.

\end{document}